\documentclass[showpacs,preprintnumbers,amsmath,amssymb]{revtex4}
\usepackage{epsfig,amssymb,euscript} 
 \def\be{\begin{equation}}
 \def\ee{\end{equation}}
 \def\bea{\begin{eqnarray}}
 \def\eea{\end{eqnarray}}

\newcommand{\beq}{\begin{equation}}
\newcommand{\eeq}{\end{equation}}
\newcommand{\beqa}{\begin{eqnarray}}
\newcommand{\eeqa}{\end{eqnarray}}
\newcommand{\beqar}{\begin{eqnarray*}}
\newcommand{\eeqar}{\end{eqnarray*}}

\newcommand{\mt}[1]{\textrm{\tiny #1}}

\def\nc {N_c}
\def\gym {g_\mt{YM}}
\def\cM{{{\cal M}_5}}

\newcommand{\reef}[1]{(\ref{#1})}

\newcommand{\eg}{{\it e.g.,}\ }
\newcommand{\ie}{{\it i.e.,}\ }

\newcommand{\norm}[1]{\raise.3ex\hbox{:}#1\raise.3ex\hbox{:}}

\newcommand\vareps{\varepsilon}

\newcommand\tC{{\widetilde C}}

\newcommand\gs{g_s} 

\newcommand\Ch{{\widehat C}}
\newcommand\Rh{{\widehat R}}

\def\tC{\tilde C}
\def\G5{G_\mt{5}}

\begin{document}
\preprint{DAMTP-2008-62,
UWO-TH-08/12}
\title{
Universal holographic hydrodynamics at finite coupling }
\author{Alex Buchel$^{1,2}$}
\email{abuchel@perimeterinstitute.ca}
\author{Robert C. Myers$^{1,3}$}
\email{rmyers@perimeterinstitute.ca}
\author{Miguel F. Paulos$^4$}
\email{m.f.paulos@damtp.cam.ac.uk}
\author{Aninda Sinha$^1$}
\email{asinha@perimeterinstitute.ca}

\affiliation{ $^1$ Perimeter Institute for Theoretical Physics,
Waterloo, Ontario N2L 2Y5, Canada
\\$^2$ Department of Applied Mathematics, University of Western Ontario, London, Ontario N6A 5B7, Canada
\\$^3$ Department of Physics and Astronomy, University of Waterloo, Waterloo, Ontario N2L 3G1, Canada
\\$^4$Department of Applied Mathematics and Theoretical Physics, Cambridge CB3 0WA, UK}
\date{\today}
\begin{abstract}
We consider thermal plasmas in a large class of superconformal gauge
theories described by a holographic dual geometry of the form
$AdS_5\times \cM$. In particular, we demonstrate that all of the
thermodynamic properties and hydrodynamic transport parameters for a
large class of superconformal gauge theories exhibit a certain
universality to leading order in the inverse 't Hooft coupling and
$1/N_c$. In particular, we show that independent of the
compactification geometry, the leading corrections are derived from
the same five-dimensional effective supergravity action supplemented
by a term quartic in the five-dimensional Weyl tensor.
\end{abstract}

\maketitle

\section{Introduction}
The gauge/gravity correspondence presents a powerful tool with which
to study strongly coupled gauge theories \cite{adscft}. One of the
most striking new insights is that the ratio of the shear viscosity
$\eta$ to the entropy density $s$ is universal with $\eta/s=1/4\pi$,
for any gauge theory with an Einstein gravity dual in the limit of
an infinite number of colours and large 't Hooft coupling, \ie
$\nc,\lambda\rightarrow\infty$ \cite{universe}. In fact, this result
has been conjectured to be a universal lower bound in nature, the
KSS bound \cite{KSS}. Corrections to this result arising for finite
$\nc$ and $\lambda$ can be calculated by taking into account higher
derivative corrections to the dual gravity action. These corrections
were first calculated for ${\cal N}=4$ super-Yang-Mills gauge theory
\cite{alex1,mps}
 \be\label{hd}
\frac{\eta}{s}=\frac 1{4\pi}\,\left[1+{15\,\zeta(3)\over
\lambda^{3/2}} +{5\over 16} {\lambda^{1/2}\over N_c^2}+{\cal
O}\left(N_c^{-3/2}e^{-\frac{8\pi^2N_c}{\lambda}}\right)\right]\,.
 \ee
Recently, it was noted \cite{universal} that the same corrections to
the ratio of $\eta/s$ appear for a certain ${\cal N}=1$
superconformal $U(\nc)\times U(\nc)$ gauge theory with bifundamental
matter \cite{igor} --- the identification of corrections requires an
appropriate interpretation of $\lambda$ and $N_c$ (see below). It
was conjectured in \cite{universal} that the leading corrections at
strong coupling to the shear-viscosity-to-entropy ratio of any
four-dimensional conformal gauge theory plasma are universal. In the
following, we prove this universality conjecture \cite{universal}
for a large class of four-dimensional superconformal gauge theories
with an $AdS_5\times \cM$ string theory dual (where $\cM$ is a
general Sasaki-Einstein manifold). In this context, the first higher
derivative corrections are well understood
\cite{R4,revR4,GreenStahn,Paulos} and appear at order $\alpha'^3$ in the
ten-dimensional type IIb supergravity action. Our approach will be
to reduce the action including the relevant higher curvature terms
down to five-dimensions and to demonstrate that the resulting
effective action is completely independent of the internal manifold
$\cM$. Hence for all of these theories, the question of determining
the effect of the higher derivative corrections to thermal
properties of the gauge theory reduces to a common problem of
studying the properties of asymptotically $AdS_5$ black hole within
a certain five-dimensional gravity action with a universal set of
$R^4$ corrections. Hence, written in terms of supergravity
expressions, the results for $\eta/s$ will be universal for the
class of theories described above. To convert the results to
variables of the dual gauge theory, care must be taken to apply the
appropriate AdS/CFT dictionary for a certain internal manifold
$\cM$. Our discussion shows that this universality of the
corrections to leading order in $1/\lambda$ and $1/\nc$ extends
beyond the ratio $\eta/s$ and that, in fact, the leading corrections
to any thermal or hydrodynamic properties of these plasmas will take
a universal form.

\section{Reduction to five dimensions}

As our starting point, we begin with a general solution of the
leading order type IIb supergravity equations which has the product
form $A_5\times \cM$. Our discussion will be general and we only
assume that $A_5$ and $\cM$ are Einstein manifolds with negative and
positive curvature, respectively. However, for our application
below, we will have in mind that $A_5$ is an asymptotically $AdS_5$
black hole. Beyond the usual choice of $S^5$ for $\cM$, the
following discussion will include any compact Sasaki-Einstein
manifold, including $L^{p,q,r}$ \cite{lpqr}, $Y^{p,q}$ \cite{ypq}
and $T^{1,1}$ as special cases. Implicitly, we also assume that the
only nontrivial fields contributing to this solution are the metric
and the Ramond-Ramond (RR) five-form. Now we wish to consider the
effects of the leading higher-derivative interactions to this
solution but in particular on the $A_5$ part of the spacetime. In
the ten-dimensional type IIb supergravity, the first nontrivial
corrections appear at order $\alpha'^3$ including the celebrated
$R^4$ interaction \cite{R4}, as well as a host of terms involving
the RR five-form (and curvatures) \cite{GreenStahn,Paulos}. However,
it can be shown that the these additional five-form terms make no
contributions to the equations of motion when working with a leading
order solution of the form $A_5\times \cM$ \cite{mps}. The
ten-dimensional action is then given by
\be S_\mt{IIb}=\frac 1{16 \pi G_\mt{10}}\int d^{10}x \sqrt{-\tilde
g}\left (\tilde R-\frac 1{4.5!} F_5^2+\alpha'^3\,
{\gs^{3/2}}\,f^{(0,0)}(\tau,\bar \tau)\, W\right )\,. \label{Action}
\ee
The pre-factor in front of the higher derivative term $W$ is a
modular form $f^{(0,0)}(\tau,\bar\tau)$ written in terms of the
usual axiodilaton field $\tau=a+i e^{-\phi}$ \cite{greengutperle}.
Recall that we assume the latter is constant in the leading
supergravity solution, \ie $e^\phi=\gs$, and so it will turn out
that the kinetic term for this field is not needed for our
discussion. Further note that self-duality constraint is imposed on
$F_5$ as an additional equation, beyond the equations of motion
derived from \reef{Action}. Now as described above, the only
relevant contribution to $W$ is fourth order in curvatures:
\bea W(\tC)&=&\tC_{ABCD}\, \tC^{EBCF}\, \tC^{AGH}{}_{E}\,
\tC^D{}_{GHF} -{1\over 4} \tC_{ABCD}\, \tC^{AB}{}_{EF}\,
\tC^{CE}{}_{GH}\, \tC^{DFGH}
 \label{C4} \eea
where $\tC$ is the Weyl tensor in ten dimensions. Above, we have
introduced the `tilde' to distinguish ten-dimensional objects, \eg
metric or Ricci scalar, from their five-dimensional counterparts
below. Similarly, our notation will be to use a `hat' to denote
quantities associated with the internal manifold, \eg $\Rh$ will be
the Ricci scalar of $\cM$. Quantities on $A_5$, the asymptotically
$AdS$ space in five dimensions, will remain unadorned. Further,
indices on the full ten-dimensional geometry, the AdS space $A_5$
and the internal manifold $\cM$ will be denoted $A,B,C,D,E,\ldots$,
$a,b,c,d,e,\dots$ and $m,n,p,q,r,\dots$, respectively.

Now we want to perform a Kaluza-Klein reduction on the $\cM$ to
construct the five-dimensional action which reproduces the gravity
equations of motion on $A_5$:
\be S_\mt{5}=\frac{1}{16 \pi G_\mt{5}}\int d^{5}x \sqrt{- g}\left
(R-\frac{12}{L^2}+\alpha'^3\, {\gs^{3/2}}\,f^{(0,0)}(\tau,\bar
\tau)\, W\right ) \label{Action5} \ee
where $L$ is the radius of curvature of $AdS_5$. Of course, the
reduction of the two-derivative terms is standard but we must take
care in reducing the higher curvature contribution $W$. First, note
that the formula for the Weyl tensor in $d$ dimensions is given by
\beq
 C_{abcd}=R_{abcd}-{2\over d-2} \left( g_{a[c}\,R_{d]b}-g_{b[c}\,R_{d]a}
 \right)+{2\over(d-2)(d-1)}\,R\,g_{a[c}\,g_{d]b}\ .
 \label{weyl}
\eeq
In the present background with the product form, $A_5\times \cM$, we
have
\bea \tilde C_{abcd}&=& C_{abcd}+10\left( g_{a [c}\, Y_{d] b}-g_{b
[c}\, Y_{d] a} \right)
+ 2 X g_{a [c}\,g_{d] b} \nonumber \\
\tilde C_{mnpq}&=& \hat C_{mnpq}+2 X \hat{g}_{m [p}\,\hat{g}_{q] n}
\nonumber \\
\tilde C_{m a n b}&=& -3 Y_{a b}\,\hat{g}_{mn}-\frac 45 X\, g_{ab}\,
\hat{g}_{mn} \eea
where we have defined
\bea
Y_{ab}&\equiv & \frac 1{24}\left(R_{ab}-{1\over 5} R\, g_{ab}\right ) \nonumber \\
X& \equiv& \frac 1{72}(R+\Rh)\,.\label{california} \eea
It will be important in what follows that $Y$ and $X$ vanish when
evaluated on the leading order supergravity solution and also that
$Y$ is traceless (in general), \ie $Y^{a}{}_{a}=0$.

Given these expressions, we have carefully evaluated \reef{C4} in
terms of $\hat C$, $C$, $Y$ and $X$. Here we will only indicate that
this straightforward but somewhat tedious exercise yields an
expression which has the schematic form:
\be \label{schem1} \tilde C^4= C^4+\hat C^4+ \hat C^3 X+ C^3 Y+C^3
X+ \mathcal O(Y^2,X^2,XY)\,, \ee
where any contributions that are quadratic or higher order in $Y$
and $X$ have been left implicit in the final term. Note that the
tensor structure of $\tC^4$ precludes the appearance of any terms
containing both $\hat C$ and $C$ together, \eg $\hat C^2\,C^2$.
However, the full expression certainly depends on the internal
geometry through the appearance $\hat C$. In particular, beyond the
$\hat C^4$ term, there are several cross terms combining $\hat C$
with the tensors $Y$ and $X$. As well as $\hat C^3 X$, the final
term in \reef{schem1} includes two further terms if the form $\hat
C^2 Y^2$ and $\hat C^2 X^2$.

Now we wish to consider how the five-dimensional equations of motion
are modified when these contributions \reef{schem1} are included in
the effective action \reef{Action5}. A key observation is that for
consistency of the expansion in $\alpha'^3$, we can evaluate the
contributions of these new terms using only the leading order
supergravity solution. In particular, this means that we can dismiss
the terms which are quadratic and higher order in $Y$ and $X$. Their
contribution to the equations of motion take the form, \eg
 \be\label{equals}
R_{ab}-\frac12 g_{ab}\,R - \frac{6}{L^2} g_{ab}\simeq\alpha'^3\left(
2X\,\frac{\delta X}{\delta g^{ab}}\,  C^2+X^2\,\frac{\delta
C^2}{\delta g^{ab}}+\cdots\right)
 \ee
However, as we observed above $X=0=Y_{ab}$ when evaluated on the
supergravity solution and hence all of these contributions, which
still contain one or more factors of $X$ or $Y$, vanish at this
order in the perturbative expansion. Note that it is this same
reasoning that allows us to ignore the appearance of $\Rh$ in the
definition of $X$ given by \reef{california}. In principle, this
quantity could depend on the details of the internal geometry,
however, the leading supergravity equations dictate that
$\Rh=20/L^2$ for any choice of $\cM$. Hence at this order, $X$
contains no information which distinguishes different internal
geometries and our discussion above only considered how $\cM$ might
modify the effective action through the appearance of $\hat C$.

Hence we are left to consider the two remaining terms involving
$\hat C$. The first of these, denoted by $\hat C^4$ in
\reef{schem1}, is simply the expression $W$ given in \reef{C4} but
now evaluated with the ten-dimensional Weyl tensor $\tC$ replaced by
$\hat C$. However, an explicit computation shows that $W(\hat C)=0$
for $\cM=L^{p,q,r}$ --- recall that the latter provide an infinite
family of explicit metrics for five-dimensional Sasaki-Einstein
manifolds \cite{lpqr}. This result can be extended to any general
Sasaki-Einstein manifold  using the fact that the latter produce a
supersymmetric background \cite{mp}. It is known \cite{GreenStahn,
Paulos} that, in supersymmetric backgrounds with only metric and
five-form fields, the full set of $\tilde C^4$ and five-form higher
derivative corrections must vanish. Further it can be shown that the
higher derivative terms involving the five-form do not contribute in
such backgrounds \cite{Paulos} and hence $W(\tilde C)$ must vanish
by itself. Now if one focuses on a supersymmetric background with
the product form $AdS_5\times\cM$, as described above, one finds
that this expression splits to yield: $W(\tilde C)\simeq W(C)+W(\hat
C)$. Again, in this relation, we have discarded the terms
proportional to $X$ and $Y$ which vanish when evaluated on the
supergravity solution. Further now, the Weyl tensor $C$ on $AdS_5$
vanishes and therefore one must have $W(\hat C)=0$ on the internal
space $\cM$.

\section{Schouten identities}

The final contribution in \reef{schem1} which could in principle
introduce some dependence of the effective action \reef{Action5} on
the internal manifold is that linear in X with a cubic contraction
of $\hat C$. This $\hat C^3 X$ term has the explicit form:
 \be\label{last}
4 X \left(2\, \hat C_{mnpq}\, \hat C^{mp}{}_{rs}\, \hat
C^{nrqs}-\hat C_{mnpq}\,\hat C^m{}_r{}^p{}_s\, \hat C^{nrqs}
\right)\,.
 \ee
We will argue that these terms vanish on using Schouten identities
 in five dimensions. Naively there exist
two independent contractions that are cubic in Weyl tensor:
 \bea
\hat C^3_{(1)}&:=& \hat C_{mnpq}\, \hat C^{mp}{}_{rs}\, \hat
C^{nrqs} \nonumber \\
\hat C^3_{(2)}&:=& \hat C_{mnpq}\,\hat C^m{}_r{}^p{}_s\, \hat
C^{nrqs}\,.
 \eea
However, in five-dimensions, these two expressions are related by a
Schouten identity:
\be\label{sch1} 2\,\hat C^3_{(1)}-\hat C^3_{(2)}=0. \ee
Such Schouten identities can be established as a vanishing
contraction of tensors which results in an attempt to antisymmetrize
over $d+1$ indices in $d$ dimensions. Here with three Weyl tensors,
one has 12 indices and one may antisymmetrize on 6 of them and then
contract with the remainder. Of course, however, the resulting
expression must vanish if it is evaluated in five dimensions. That
is, up to an overall normalization, the left-hand side of
\reef{sch1} is equivalent to
\be\label{shout} \hat C_{[mn}{}^{mn}\,\hat C_{pq}{}^{pq}\,\hat
C_{rs]}{}^{rs} \ee
which again vanishes in five dimensions. These calculations are
quickly performed using {\it Cadabra} \cite{cadabra}.

At this point, we have actually done enough to establish that the
five-dimensional equations of motion are independent of the internal
manifold $\cM$. However, we proceed further here with the
application of Schouten identities to eliminate the $C^3X$, $C^3Y$
terms in \reef{schem1}, as well. The latter terms work out to be
 \bea
&&4 X \left(2 C_{a b c d}\,C^{a c}{}_{e f}\,C^{b e d f}-C_{abcd}\,
C^a{}_e{}^c{}_f \,C^{bedf}\right)\nonumber\\
&&\qquad\qquad+40 Y_g{}^f\, \left(2C_{abcd}\,C^{ac}{}_{e f}\,
C^{bedg}-C_{abcd}\,C^a{}_{e}{}^c{}_f\,
C^{bedg}-C_{abcd}\,C^{acbe}\,C^d{}_{f e}{}^g\right)\,.
 \label{XY}\eea
Now the first two terms proportional to $X$ again cancel by the same
Schouten identity given in \reef{sch1}. Using a similar strategy as
above, it is possible to show that even though in principle one can
build three independent $C^3Y$ contractions, namely,
\bea
(C^3Y)_{(1)}&:=&C_{abcd}\,C^{ac}{}_{e f}\,C^{bedg}\, Y_g{}^{f} \nonumber \\
(C^3Y)_{(2)}&:=&C_{abcd}\,C^a{}_e{}^c{}_f\, C^{bedg} \,Y_g{}^{f} \nonumber \\
(C^3Y)_{(3)}&:=&C_{abcd}\,C^{acbe}\,C^{d}{}_{fe}{}^{g}\, Y_g{}^{f}
 \nonumber
 \eea
there is now a new Schouten identity
 \be \label{sch2}
2 (C^3Y)_{(1)}- (C^3Y)_{(2)}- (C^3Y)_{(3)}=0\,.
 \ee
To obtain this identity, one can take for instance
 \be
Y_{[a}{}^{f}\,C_{bc}{}^{ad}\,C_{d}{}^{beg}\,C_{ef]g}{}^{c}
 \ee
which must again vanish in five dimensions. In any event, this new
identity \reef{sch2} ensures that the combination of $C^3Y$ terms in
\reef{XY} vanishes.

\section{Universal corrections}

Our discussion above shows that the five-dimensional Einstein
equations on the asymptotically AdS space $A_5$ will not depend on
the detailed structure of the compact manifold $\cM$. In fact with
the Schouten identities, we were able to show that upon reducing to
five dimensions the quartic curvature term in \reef{Action5} reduces
to
\be \label{schem7} W(\tC)= W(C)+ \mathcal O(Y^2,X^2,XY) \ee
where $W(C)$ is the expression \reef{C4} constructed with the
five-dimensional Weyl tensor. Further, as discussed at
\reef{equals}, the terms which are quadratic or higher order in $X$
and $Y$ will not contribute to the equations of motion at this order
in the $\alpha'$ expansion. That is, the $\alpha'$ corrected
equations for the five-dimensional metric give the same result
whether one uses the full $W(\tC)$ or simply treats this expression
a five-dimensional construction $W(C)$.

Of course, we must now consider the implications of this result for
the dual gauge theory. In particular, we are interested in the
thermodynamic and hydrodynamic properties of the gauge theory. In
this case, we will take $A_5$ to be an asymptotically $AdS_5$ black
hole. The result of our above discussion is that for all of the
gauge theories (defined by different $\cM$), the dual
five-dimensional gravity action is universal. Therefore determining
the effect of the higher derivative gravity corrections reduces to a
common problem of studying the properties of asymptotically $AdS_5$
black holes within the same five-dimensional effective theory with a
universal set of $R^4$ corrections. Hence the corrections to all of
the thermodynamic properties (\eg entropy density or equation of
state) and hydrodynamic parameters (\eg shear viscosity or
relaxation time) of the gauge theories will have a universal form.
In particular then, we have proven the universality of corrections
to the quasinormal spectrum, as conjectured in \cite{universal}. Of
course, the latter spectrum captures a great deal of information
about the thermal transport coefficients. However, we emphasise that
our result extends this universal feature of the $\alpha'^3$
corrections to the full set of higher order coefficients recently
explored in \cite{hydro}, including those for terms nonlinear in the
local four-velocity. Again our present discussion indicates that it
is sufficient to consider the effective action \reef{Action5} with
$W$ term constructed with five-dimensional Weyl tensors to calculate
the $\alpha'^3$ corrections to all of these transport coefficients.

We must add that the above conclusions rely on the physical
quantities of interest not being modified by new fields which are
trivial in the original background. For example, we know that
although the dilaton and the warp factor are trivial in the
$AdS_5$ black hole solution of the leading order supergravity
equations, both of these fields are sourced by the higher curvature
corrections in this background \cite{GKT,PT2,alex1}. Further, our
analysis is restricted to the $\alpha'^3$ terms in the
ten-dimensional action, which include only the curvatures and the RR
five-form \cite{GreenStahn,Paulos}. However, we know that there
exist a host of additional higher derivative interactions at order
$\alpha'^3$ \cite{pure} and in principle, even more additional type
IIb fields could be sourced by these terms. For definiteness,
consider the RR axion $a$ which vanishes at lowest order. There
might still be $\alpha'^3$ terms which are linear in $a$, \eg $ C^2
\nabla F^+_5 \nabla^2 a$. The corrected solution would then also
include an axion of order $\alpha'^3$. However, in five-dimensional
Einstein's equations, $a$ will only appear quadratically or in terms
with an $\alpha'^3$ factor. Hence, its effects in, \eg the
quasi-normal spectrum will only be felt at order $\alpha'^6$ and
therefore it can be neglected here. The same reasoning can be made
for all other fields, including the dilaton and warp factor.

At this point, we have reduced the determination of higher order
corrections in a large number of CFT's down to the study of a common
five-dimensional gravity theory. The latter gives a universal set of
corrections in terms of supergravity expressions. As an example, let
us consider the ratio of shear viscosity to entropy density:
 \be
\frac{\eta}{s}=\frac{1}{4\pi}\left[1+15\frac{\alpha'^3}{L^6}\left(\zeta(3)+
\frac{\pi^2}{3}\gs^2+ \tilde{f}_\mt{NP}\right)\right]\,.
 \label{walk1}
 \ee
This result is produced by expanding the modular formula
$f^{(0,0)}(\tau,\bar \tau)$ in a regime of small string coupling
$\gs$ \cite{greengutperle}. As \reef{walk1} shows, in this regime,
the modular form can be interpreted in terms of a tree-level term, a
one string-loop contribution and a series of nonperturbative
corrections captured by $\tilde{f}_\mt{NP}$. Again with small $\gs$,
the leading nonperturbative contribution is $\tilde{f}_\mt{NP}
\approx 4\pi \gs^{3/2} e^{-2\pi/\gs}$ \cite{comm}. Now in the case
of ${\cal N}=4$ SYM for which $\cM=S^5$, we have the standard
AdS/CFT dictionary which includes $\lambda=L^4/\alpha'^2$ and
$\gym^2=\lambda/\nc=4\pi\gs$. Applying these two expressions
converts the above supergravity expression \reef{walk1} to the dual
gauge theory expression \reef{hd} given in the introduction.

Now in the case where $\cM$ is replaced by a more general
Sasaki-Einstein manifold, one must take care in interpreting
\reef{walk1} with the correct AdS/CFT dictionary. In this case, the
dual gauge theory corresponds to a quiver theory containing a number
of $U(\nc)$ gauge groups coupled to certain bifundamental matter
fields \cite{mp}.  Hence with a product of $n$ gauge
groups, there are $n$ independent gauge couplings $(\gym^2)_i$. The
string coupling is related to the following combination \cite{mp}

 \be
\frac{1}{\gs}=\sum_i\frac{4\pi}{(\gym^2)_i}\,,
 \label{dictate1}
 \ee
while the remaining independent linear combinations of $(\gym^2)_i$
are related to various form fields on $\cM$. It will prove useful to
define a ``collective'' 't Hooft coupling for the quiver theory with
an averaged gauge coupling:
 \be
\lambda_\mt{CFT}\equiv\bar{g}_\mt{YM}^2\nc\qquad{\rm where}\ \
\frac{1}{\bar{g}_\mt{YM}^2}\equiv\frac{1}{n}\sum_i\frac{1}{(\gym^2)_i}\,.
 \label{dicta0}
 \ee
Now the AdS radius of
curvature on $A_5$ is determined by \cite{Gubser4}
 \be
\frac{L^4}{\alpha'^2}=4\pi\,\gs\,\nc\, \frac{\pi^3}{{\rm
Vol}(\cM)}\,, \label{dic0}
 \ee
where the internal volume is defined for $\cM$ with unit curvature,
\ie with $\Rh=20$. Further the volume of the internal space is
related to the central charge of the dual quiver theory as
\cite{Gubser4}
 \be
\frac{c_\mt{CFT}}{c_{\scriptscriptstyle {\cal
N}=4}}=\frac{\pi^3}{{\rm Vol}(\cM)}\label{dict0}
 \ee
where $c_{\scriptscriptstyle {\cal N}=4}=(\nc^2-1)/4$ is the central
charge of ${\cal N}=4$ super-Yang-Mills with gauge group $SU(\nc)$.
Combining these last three equations then yields
 \be\label{dictate2}
\frac{L^4}{\alpha'^2}=\frac{\lambda_\mt{CFT}}{n}
\,\frac{c_\mt{CFT}}{c_{\scriptscriptstyle {\cal N}=4}}\,.
 \ee
 Combining these above expressions
allows us to translate \reef{walk1} to the following gauge theory
result
 \be
\frac{\eta}{s}=\frac{1}{4\pi} \left[
1+\left(\frac{n\,c_{\scriptscriptstyle {\cal
N}=4}}{c_\mt{CFT}}\right)^{3/2}\frac{15\,\zeta(3)}{\lambda_\mt{CFT}^{3/2}}+
\left(\frac{n\,c_{\scriptscriptstyle {\cal
N}=4}}{c_\mt{CFT}}\right)^{3/2}
\frac{5}{16n^2}\frac{\lambda_\mt{CFT}^{1/2}}{\nc^2} +{\cal O}
\left( N_c^{-3/2} e^{- 8\pi^2n
\frac{N_c}{\lambda_\mt{CFT}}}\right)\right]\,.
 \label{wok0}
 \ee
 As we can see while the basic structure of
the corrections in \reef{hd} and \reef{wok0} are the same, the
precise numerical coefficients differ in a way dictated by the
distinct gauge/gravity duality for each choice of the internal
geometry. As an illustrative example, let us consider the case where
$\cM=T^{1,1}$. With this internal space, the dual quiver theory has
$n=2$ nodes and corresponds to an ${\cal N}=1$ superconformal
$U(\nc)\times U(\nc)$ gauge theory coupled to four chiral
superfields, two each in the $(\nc,\bar{\nc})$ and $(\bar{\nc},\nc)$
representations, as originally elucidated by Klebanov and Witten
\cite{igor}. Further ${\rm Vol}(\cM)=16\pi^3/27$  and so for the Klebanov-Witten theory,
\reef{wok0} yields
 \be
\left.\frac{\eta}{s}\right|_\mt{KW}=\frac{1}{4\pi} \left[
1+\left(\frac{32}{27}\right)^{3/2}\frac{15\,\zeta(3)}{\lambda_\mt{KW}^{3/2}}+
\left(\frac{32}{27}\right)^{3/2}\frac{5}{64}\frac{\lambda_\mt{KW}^{1/2}}{\nc^2}
+{\cal O} \left( N_c^{-3/2}
e^{-\frac{16\pi^2N_c}{\lambda_\mt{KW}}}\right)\right]\,.
 \label{wok1}
 \ee

\section{Discussion}

In this note, we have extended the universality of $\eta/s$ to the
regime of a large but finite 't Hooft coupling and finite number of
colours for the class of superconformal field theories described by
a holographic dual geometry of the form $AdS_5\times\cM$. Our proof
followed by constructing the effective five-dimensional gravity
action which results from compactifying with type IIb supergravity
action supplemented by the $\tC^4$ term \reef{C4} in 10 dimensions
on some geometry $\cM$. We demonstrated that the resulting
five-dimensional equations of motion are independent of $\cM$ and in
fact, the latter equations can be derived from an effective action
with the same quartic curvature term \reef{C4} constructed from the
five-dimensional Weyl tensor. This shows that the leading
corrections to $\eta/s$ are derived for a large class of theories by
studying the same problem, \ie determining the effect of the higher
derivative gravity corrections to an asymptotically $AdS_5$ black
hole within the same five-dimensional effective action with a
universal set of $C^4$ corrections. Hence, written in terms of
string theory variables (string tension and string coupling), the
result \reef{walk1} for $\eta/s$ will be universal for a large class
of theories. To convert the result to variables of the dual gauge
theory, one must apply the AdS/CFT dictionary that is appropriate
for a given internal manifold $\cM$. This universality was first
conjectured for the corrections to the quasinormal spectrum of the
dual $AdS_5$ black holes \cite{universal}. Our proof of universality
encompasses this conjecture but also shows that this universal
behaviour extends to all other thermodynamic or hydrodynamic
quantities \eg the entropy density \cite{GKT} and the relaxation
time \cite{relax}, in the same way as described above.

One step in our analysis was to show the quartic curvature term
\reef{C4} vanishes when evaluated for the internal Weyl tensor $\hat
C$. We presented a general argument that relied on supersymmetry and
so required choosing $\cM$ to be a five-dimensional Sasaki-Einstein
space. It would be interesting to generalise this discussion beyond
supersymmetric compactifications. One infinite family of Einstein
spaces can be constructed as coset spaces $(SU(2)\times
SU(2))/U(1)$. The resulting manifolds, denoted $T^{p,q}$, have been
considered for type IIb compactifications in
\cite{Romans,Gubser4,Gubser5}. Of these only $T^{1,1}$ provides a
supersymmetric background \cite{Romans} and in fact, with $p\ne q$,
the compactifications contain tachyons violating the
Breitenlohner-Freedman bound \cite{Gubser5}. In any event,
evaluating $W(\hat C)$ for these spaces, we find that it only
vanishes if $p=q$. Even though the case $p=q>1$ is not
supersymmetric, one should expect the quartic curvature term to
vanish on these spaces since it vanishes on the supersymmetric
$T^{1,1}$ geometry and the spaces $T^{p,p}$ are {\bf Z}$_p$
orbifolds of $T^{1,1}$. Hence it seems that the vanishing of $W(\hat
C)$ is closely related to supersymmetry but it would be interesting
to examine this question further.

Another interesting direction would be to see if our findings extend
to the case of non-zero chemical potential. In this case, the
gravity analysis would require turning on background Kaluza-Klein
gauge fields and so the resulting background would not have the
product form $A_5\times \cM$ assumed throughout the above
discussion. In particular, we expect that the higher derivative
terms involving the RR five-form will play a crucial role
\cite{mps}. Further, different choices of $\cM$ will in general lead
to different gauge fields in the effective five-dimensional action
and so we cannot expect universality to extend to any general
chemical potential in the dual gauge theory. However, it was
recently shown that any supersymmetric compactification on a
Sasaki-Einstein manifold yields a consistent reduction to minimal
five-dimensional supergravity \cite{jerome}. This construction
relies on using the Killing symmetry of $\cM$ which is dual to the
$U(1)$ R-symmetry in the dual ${\cal N}=1$ superconformal gauge
theory. Hence, with an infinite 't Hooft coupling and an infinite
number of colours, the thermal and hydrodynamic properties of the
gauge theory plasma will be universal in the presence of this
corresponding chemical potential. It is likely that this
universality could also be extended to finite $\lambda$ and $\nc$ in
this case.

Above we observed that even though the $\alpha'^3$ corrections to
the supergravity action may source a nontrivial dilaton, warp factor
and other new fields, these new fields will not modify the thermal
and hydrodynamic properties of the dual gauge theory at this order.
The appearance of additional fields at the $\alpha'^3$ order would
indicate that various operators acquire a nonvanishing expectation
value in the gauge theory plasma, but that these expectation values
are suppressed by inverse powers of the 't Hooft coupling and the
number of colours. While certain fields, \eg the dilaton, may
exhibit universal behaviour, this could not be expected to apply for
all of the new operators acquiring an expectation value. In
particular, the precise family of operators available would
certainly depend on the details of the gauge theory or alternatively
the compactification manifold.

The warp factor deserves further attention as we would like to
relate our conclusions to the discussion in \cite{universal}.
Specifically, it was shown in \cite{universal} that the equation of
motion for the shear quasinormal mode obtained in ten-dimensional
using $\tilde C^4$ higher derivative metric corrections differed
from the corresponding equation derived in five-dimensional
effective action with $C^4$ term only. While the wave-functions of
the  lowest (hydrodynamic) shear quasinormal modes were different,
hydrodynamic shear dispersion relation was found to be the same
\cite{universal}. We point out here that a simple ${\cal
O}(\alpha'^3)$ rescaling of the five-dimensional shear quasinormal
wave-function $Z_{shear,5D}$
 \be\label{5d10d}
Z_{shear,5D}\to Z_{shear,5D}\cdot\left(1-{3\over 10}\alpha'^3
\nu\right)\equiv Z_{shear,10D}
 \ee
identified it with the ten-dimensional shear quasinormal
wave-function $Z_{shear,10D}$. In \reef{5d10d} $\nu$ is a warp
factor modifying the direct product $A_5\times \cM$ due to higher
derivative corrections \cite{GKT}:
 \be\label{metdiff}
ds_{A_5}^2\times ds_{\cM}^2 \to \left\{ds_{A_5}^2\cdot  e^{-{10\over
3}\alpha'^3\nu}\right\}\times \left\{ds_{\cM}^2 \cdot e^{2\alpha'^3
\nu}\right\}
 \ee
Since the warp factor $\nu$ is non-singular (and momenta
independent), a rescaling \reef{5d10d} changes the quasinormal
wave-function while keeping the spectrum of the quasinormal modes
invariant. Such a rescaling is simply an artifact of the (arbitrary)
normalization used in \cite{bb} (and later in \cite{universal}) ---
a ten-dimensional shear quasinormal wave-function was normalized as
 \be
Z_{shear,10D}\propto g^{xx}_{5D}\cdot \delta g
 \ee
where $\delta g$ is the corresponding shear metric fluctuation. If
instead  the wave-functions are normalized as
 \be
Z\propto g^{xx}\cdot \delta g
 \ee
in five  and ten dimensions, correspondingly, there is no need for a
rescaling \reef{5d10d}.

As displayed in \reef{schem7}, the ten-dimensional quartic curvature
corrections reduce down to $W(C)$ constructed from the
five-dimensional Weyl tensor, plus a number of terms that are
quadratic or higher order in $X,Y$.  While these terms only
contribute to the equations of motion with expressions that vanish
when evaluated on the leading order supergravity solution, in
appendix \ref{remove}, we show that we can use field redefinitions
to remove these terms and so the resulting equations of motion would
identical independent of the choice of $\cM$. However, we note that
such field redefinitions would make it difficult to realize the
supersymmetry of the various $AdS_5\times\cM$ backgrounds. Of
course, the latter is not central to the present investigation but
remarkably we were able to demonstrate universality without
resorting to such field redefinitions. Note that this would not have
been possible if the terms linear in $X$ and $Y$ had not vanished by
the Schouten identities.


Our analysis has demonstrated that the thermal properties of a large
class of gauge theories can be derived from a universal holographic
framework. Of course, one is tempted to consider the application of
our results to the strongly coupled quark-gluon plasma under study
in experiments at RHIC and soon at the LHC. A key difference between
${\cal N}=4$ SYM and QCD is the number of degrees of freedom that
are active in the strongly coupled plasmas \cite{compare,compare2}.
Our analysis captures the effect of the latter on thermal properties
and, in particular, on the $1/\nc,\ 1/\lambda$ corrections with the
dependence on the central charge, \eg with the factors proportional
to $c_{\scriptscriptstyle {\cal N}=4}/c_\mt{CFT}$ in \reef{wok0}.
Imagining that the QCD plasma is described by a CFT, we can proceed
by treating its central charge to be a phenomenological parameter.
Let us then consider the energy density and shear viscosity arising
from the present holographic model:
 \be
\frac{\vareps}{\vareps_0}=\frac{3}{4}\left(1 +
\frac{\Delta}{8}\right) \qquad{\rm and}\qquad
\frac{\eta}{s}=\frac{1}{4\pi}\left(1+\Delta \right) \,.
 \label{new1}
 \ee
 \be
{\rm where}\qquad \Delta\equiv5\left(\frac{c_{\scriptscriptstyle
{\cal
N}=4}}{c_\mt{QCD}}\right)^{3/2}\left(\frac{3\,\zeta(3)}{\lambda^{3/2}}+
\frac{1}{16}\frac{\lambda^{1/2}}{\nc^2}\right)\,.
 \label{new2}
 \ee
Here $\vareps$ and $\vareps_0$ denote the energy density of the
conformal plasma and that in limit of a free theory. We have set
$n=1$ in \reef{new2} and recall that $c_{\scriptscriptstyle {\cal
N}=4}=2$ with an $SU(3)$ gauge group. Lattice QCD results can
provide insight into the energy density and recent studies seem to
indicate that energy density should be in the range
$\vareps/\vareps_0\approx0.85\,-\,0.90$ \cite{lattice}. In this
case, \reef{new1} yields $\Delta\approx1.07\,-1.60$ and hence
 \be
\left.\frac{\eta}{s}\right|_\mt{QCD}\approx0.16\,-\,0.21\ .
 \label{fantasy}
 \ee
We must observe that the corrections here are not small. These
`corrected' values for $\eta/s$ are significantly larger than
leading result, which corresponds to the conjectured KSS bound
$\left.\eta/s\right|_\mt{KSS} =1/4\pi \simeq.08$ \cite{KSS}.
However, these results \reef{fantasy} are still consistent with
values emerging from the analysis of RHIC data \cite{rick}. Of
course, our analysis must be regarded with a highly skeptical eye.
We are assuming that that the QCD plasma is described by a
supersymmetric conformal field theory. In fact for the QCD plasma,
we certainly have no supersymmetry and at RHIC temperatures, we
should expect that it is only approximately conformal.

Another interesting comparison can be made for pure $SU(3)$
Yang-Mills, again using lattice results \cite{thanks}. While lattice
results for the thermodynamics of the pure gauge theory have long
been available \cite{yangmills}, reliable results for $\eta/s$ have
only been established very recently \cite{meyer}:
 \be
 \eta/s\approx
  0.10\,-\,0.17\quad {\rm at}\ T=1.65\,T_c\,.
 \label{one1}
 \ee
Examining the thermodynamic results \cite{yangmills} at this
temperature and applying the same analysis as above, one finds
 \be
  T=1.65\,T_c\,:\ \vareps/\vareps_0\approx0.81\,-\,0.84\,,\
  \Delta\approx0.64\,-\,0.96\,,\ \eta/s\approx0.13\,-\,0.16\,.
 \label{two2}
 \ee
Hence the holographic formulae yield results in good agreement with
those from the lattice \reef{one1} --- however, the errors are still
relatively large. We should mention that \cite{meyer} also presents
a result for $T=1.24\,T_c$. However, this close to the critical
temperature, one finds that $\vareps/\vareps_0\approx0.71\,-\,0.74$.
Achieving $\vareps/\vareps_0<3/4$ in \reef{new1} would require
$\Delta<0$ and so one cannot apply the present holographic model
because our formula \reef{new2} always gives $\Delta>0$. Of course,
one should also note that the lattice results indicate that the
interaction measure, \ie $(\varepsilon-3p)/T^4$, peaks just below
$T=1.24\,T_c$ \cite{yangmills} and so it seems the Yang-Mills
plasma is out of the conformal regime at this temperature. For
comparison purposes, we also apply our calculations for $T=4\,T_c$,
where the plasma appears to be well into the conformal regime:
 \be
  T=4\,T_c\,:\ \vareps/\vareps_0\approx0.83\,-\,0.86\,,\
  \Delta\approx0.88\,-\,1.20\,,\ \eta/s\approx0.15\,-\,0.17\,.
 \label{three3}
 \ee
Of course, just as for the QCD plasma, this holographic analysis
should be considered in a skeptical light.

Note that the results in \reef{fantasy}, \reef{two2} and
\reef{three3} rely simply on the form in \reef{new1} arising from
the holographic model and are actually independent of any
microscopic details \ie $\lambda$, $\nc$ or ${c}_\mt{QCD}$. If we
adopt $\lambda=6\pi$ (\ie $\alpha_s=0.5$) along with $\nc=3$ for the
QCD plasma, we may use \reef{new2} to derive the effective central
charge: ${c}_\mt{QCD}\approx 0.75\,-\,0.90$. A general observation
is that this reduced central charge (compared to ${\cal N}=4$ SYM)
should reflect the reduction in degrees of freedom for QCD and is
responsible for the enhancement of the the $1/\nc,\ 1/\lambda$
corrections. For example, with the same 't Hooft coupling, the
corrected result for ${\cal N}=4$ SYM in \reef{hd} yields
$\left.\eta/s\right|_{\scriptscriptstyle {\cal N}=4}\approx0.11$
\cite{mps}.

We may also calculate the effective central charge for the pure
Yang-Mills plasma at $T=1.65\,T_c$ or $4\,T_c$. This requires that
we also use the lattice results \cite{yangmills} for the Yang-Mills coupling at these
termperatures, which yields: $\lambda_{1.65T_c} \approx
6.00\,-\,6.35$ and $\lambda_{4T_c} \approx 6.61\,-\,7.05$. With
these couplings, we find the effective central charge to be:
$c_{1.65T_c}\approx 2.34\,-\,3.23$ and $c_{4T_c}\approx
1.84\,-\,2.39$. This is somewhat surprising since here this
effective central charge for the pure gauge theory is of the same
order as that for the ${\cal N}=4$ SYM, \ie $c_{\scriptscriptstyle
{\cal N}=4}=2$ but one's intuition would be that the SYM plasma
should contain more degrees of freedom than that for the pure gauge
theory. In any event, these results emphasize that $c$ is only a
phenomenological parameter in our holographic model.

Finally we close by observing that four-dimensional supersymmetric
CFT's are characterised by two central charges $a$ and $c$ defined
in terms of the trace anomaly --- see \cite{bng} for a detailed
discussion in the context of gauge/gravity correspondence. All of
the AdS/CFT dualities considered here were based on type IIb
compactifications on a smooth five-dimensional manifold $\cM$ and
the leading higher derivative corrections appeared at order
$\alpha'^3$. Hence an implicit feature common to all of the dual
conformal gauge theories is that the $a$ and $c$ central charges are
equal. Of course, it is possible to construct AdS/CFT dualities in
which $a\ne c$ but such F-theory constructions typically require the
introduction of various D-branes and O-planes \cite{ofer}. Hence a
detailed analysis of such ten-dimensional constructions would
require more than just using the type IIb supergravity action, as
was done above. A key feature of the resulting five-dimensional
effective action is that the leading corrections are now
curvature-squared terms with a coefficient proportional to $a-c$
\cite{bng,others}. Therefore such theories are beyond the scope of
the present analysis and the universality of the shear viscosity (or
any other thermal properties) found here does not apply to such
conformal gauge theory plasmas with $a\ne c$. In fact, these
theories can violate the conjectured KSS bound \cite{violate}, in
contrast to the present case where $1/\lambda$ and $1/\nc$
corrections are always positive and so still respect the KSS bound,
as has been noted in \cite{alex1,mps}.

\acknowledgments{We thank Jerome Gauntlett, Jaume Gomis, Krishna
Rajagopal and Andrei Starinets for useful discussions. Further we
thank Kasper Peeters for his wonderful program {\it Cadabra} and for
very useful technical help with it. Research at Perimeter Institute
is supported by the Government of Canada through Industry Canada and
by the Province of Ontario through the Ministry of Research \&
Innovation. AB gratefully acknowledges further support by an NSERC
Discovery grant and support through the Early Researcher Award
program by the Province of Ontario. RCM also acknowledges support
from an NSERC Discovery grant and funding from the Canadian
Institute for Advanced Research. MP is supported by the Portuguese
Fundacao para a Ciencia e Tecnologia, grant SFRH/BD/23438/2005. MP
also gratefully acknowledges Perimeter Institute for its hospitality
at the beginning of this project.

\begin{appendix}

\section{Field redefinitions} \label{remove}
Here we will show that using field redefinitions all of the terms
proportional to $X$ or $Y$ in (\ref{schem1}) can be set to zero.
Suppose the effective five-dimensional gravity action contains two
terms
 \beq
 M^{ab}\,Y_{ab}+ N\,X
 \label{newer}
 \eeq
where $M_{ab}$ and $N$ are some functionals of the five-dimensional
curvature (and perhaps other fields). Now we will show that these
two terms can be eliminated by a field redefinition. First of all
note that in the supergravity background $\Rh=20/L^2$ and therefore
we can write the leading five-dimensional supergravity action as
 \beq
 I_{sugra}={1\over 16\pi \G5}\int d^5x\,\sqrt{-g}\,\left[R+{12 \over
 L^2}\right]={1\over 16\pi \G5}\int d^5x\,\sqrt{-g}\,\left[R+{3 \over
 5}\Rh\right]\ .
 \label{hockey}
 \eeq
Now considering a field redefinition of the five-dimensional metric,
$g_{ab}\rightarrow g_{ab}+\delta g_{ab}$, the change in the
five-dimensional supergravity action is (up to total derivatives):
 \beqa
 \delta I_{sugra}&=&{1\over 16\pi \G5}\int d^5x\,\sqrt{-g}\,\left[-R^{ab}+{1\over 2}
 \left(R+{3 \over 5}\Rh\right)g^{ab}\right]\,\delta g_{ab}
 \nonumber\\
 &=&{1\over 16\pi \G5}\int d^5x\,\sqrt{-g}\,\left[-\left(R^{ab}-{R\over 5}g^{ab}\right)
 +{1\over 2}\left(-{2\over5}R+R+{3 \over 5}\Rh\right)g^{ab}\right]\,\delta g_{ab}
 \nonumber\\
 &=&{1\over 16\pi \G5}\int d^5x\,\sqrt{-g}\,\left[-24\,Y^{ab}
 +{108\over 5}X\,g^{ab}\right]\,\delta g_{ab}\ .
 \label{step1}
  \eeqa
Now making the specific choice
 \beq
 \delta g_{ab}=\alpha\,M_{ab}+\left(\beta\,M^c{}_c+\gamma\,N\right)\,g_{ab}
 \label{step2}
 \eeq
then \reef{step1} yields
 \beqa
 \delta I_{sugra} &=&{1\over 16\pi \G5}\int d^5x\,\sqrt{-g}\,\left[-24\alpha\,M_{ab}\,Y^{ab}
 +108\left(\beta+{\alpha\over 5}\right)\,X\,M^c{}_c+108\gamma\,N\,X\right]
 \nonumber\\
 &=&{1\over 16\pi \G5}\int d^5x\,\sqrt{-g}\,\left[-M_{ab}\,Y^{ab}
 -N\,X\right]
 \label{step3}
 \eeqa
where in the last line, we chose $\alpha=1/24$, $\beta=-\alpha/5$
and $\gamma=-1/108$. Hence by tuning the parameters appropriately we
can cancel the extra order $\alpha'^3$ terms \reef{newer} from the
effective five-dimensional action. Note that this is quite general
and these calculations show that one can eliminate any term
containing an $X$ or $Y$ -- in particular, such a term need not be
linear in $X$ and $Y$. Hence all the higher order terms found above
in \reef{schem1} can in principle be completely eliminated leaving
only the $C^4$ and $\Ch^4$ terms. Hence one is lead to conclude that
a field redefinition can be used to bring the five- and
ten-dimensional equations into precisely the same form, \ie not just
the same up to terms which vanish when evaluated for the leading
order supergravity solution.

\end{appendix}


\begin{thebibliography}{99}
\bibitem{comm}
Implicitly we have set the RR axion to zero here. A nonvanishing
axion would only appear here in the nonperturbative contributions.
\end{thebibliography}
\end{document}